# Patterns of Selection of Human Movements II: Movement Limits, Mechanical Energy, and Very Slow Walking Gaits

**Stuart Hagler**
Oregon Health & Science University
Portland, OR, USA
haglers@ohsu.edu

**Abstract:** The biomechanics of the human body allow humans a range of possible ways of executing movements to attain specific goals. This range of movement is limited by a number of mechanical, biomechanical, or cognitive constraints. Shifts in these limits result in changes available possible movements from which a subject can select and can affect which movements a subject selects. Therefore by understanding the limits on the range of movement we can come to a better understanding of declines in movement performance due to disease or aging. In this project, we look at how models for the limits on the range of movement can be derived in a principled manner from a model of the movement. Using the example of normal walking gaits, we develop a lower limit on the avg. walking speed by examining the process by which the body restores mechanical energy lost during walking, and we develop an upper limit on the avg. step length by examining the forces the body can exert doing external mechanical work, in this case, pulling a cart. Making slight changes to the model for normal walking gaits, we develop a model of very slow walking gaits with avg. walking speeds below the lower limit on normal walking gaits but that also has a lower limit on the avg. walking speed. We note that the lowest avg. walking speeds observed clinically fall into the range of very slow walking gaits so defined and argue that forms of bipedal locomotion with still lower speeds should be considered distinct from walking gaits.

1 Introduction

Due to the complexity of human biomechanics (e.g. articulated limbs), the execution of most human movements requires sophisticated, coordinated sensory-motor and cognitive processes. [1, 2] Subjects select movements from a range of movement that is limited by a number constraints: mechanical (e.g. limits on the poses the human body can take), biomechanical (e.g. limits on the forces muscles can generate), and cognitive (e.g. limits on how much information the brain can process while the movement is going on; cf. the speed/accuracy trade-off for rapid, targeted movements described by Fitts' law [3-5]). Shifts in these limits result in changes in the range from which a subject selects a movement which in turn can potentially affect the movements the subjects select. The goal of this project is to investigate the principles that limit the range of human movements. The general idea is that models of limits can be developed out of the model of the movement by placing constraints on mechanical, biomechanical, or cognitive processes in the model.

For an example of how we might study declines in movement performance by modeling limits on the range of movements, we can look at walking gaits. Healthy adults generally exhibit a robust walking gait with avg. walking speed about $1.3 \text{ m} \cdot \text{s}^{-1}$ and avg. step length about 0.61 m. [6] However, subject's diagnosed with disorders such as Parkinson's disease (PD), Multiple Sclerosis, or Alzheimer's disease often exhibit a much slower walking gait with walking speeds of about $0.5 \text{ m} \cdot \text{s}^{-1}$ to $0.6 \text{ m} \cdot \text{s}^{-1}$ and avg. step lengths of about 0.42 m to 0.50 m. [7] As we indicated in [8], evidence supports the use of avg. walking speed tests as predictors of adverse results related with health in older adults, [9] and has shown avg. walking speed to be a quantitative estimate of risk of future hospitalization. [10] Studies have shown a relationship between avg. walking speed and cognition. [11, 12] Slower avg. walking speed has been demonstrated in dementia patients compared to controls, [13] and has been shown to precede cognitive impairment [14] and dementia, [15] and timed walk has been used as a partial characterization of lower extremity function which has been shown to predict disability. [16, 17] The slowing of avg. walking speed appears to take place secondary to the slowing of processing speed in the path leading to dementia. [18] Declines in walking gait performance generally include decreases in both the avg. walking speed and avg. step length. As these two values are, in general, highly correlated, [19] the decline in performance can be due to a downward shift on an upper limit on the avg. walking speeds or a downward shift on an upper limit on the avg. step lengths. The limits denote where some process required for the execution of the movement breaks down in such a way that the movement can no longer be carried out. In so far as it is these processes that we are using the movement as a vehicle to study (e.g. using observations of declines in walking gait performance to study cognitive declines in older adults), then the limits, and models of the limits, should provide a clearer picture of the process of interest.

The approach we take to modeling the limits of a range of movement is to begin with a model of the movement of interest and to construct limits using the model by exploring the areas where the model breaks down; we use walking gait as an example for how this approach works in practice. We carry out this project out in four parts. In the first part (Sec. 2), we provide an outline of the formalism we use to describe the metabolic and mechanical energies, paying particular attention to the concepts of external mechanical work and mechanical energy loss which are important concepts in the subsequent sections. In the second part (Sec. 3), we construct a model for normal walking gaits doing external mechanical work that we use as the basis for the construction of limits on the avg. walking speed and avg. step length in the following sections. We show that the model for normal walking gaits doing external work provides an account of the empirical data for walking while pulling a cart reported by Atzler & Herbst, [20] In the third part (Sec. 4), we estimate the mechanical energy loss during normal walking gaits doing external mechanical and obtain two lower limits on the avg. walking speeds of walking gaits. We define the range between these two limits to be that of very slow walking gaits and observe that this range of very slow walking gaits contains the slowest walking gaits for younger adults reported by Grieve, [19] for older adults reported by Hagler et al., [21] and for pre-rehabilitation PD subjects reported by Frazzitta et al. [7] In the fourth part (Sec. 5), we use the model developed in Sec. 3 to estimate the forces the subject must generate to execute normal walking gaits doing external work. We find that by placing a limit on the maximum forces the subject can generate we obtain an upper limit on the avg. step length when the normal walking gait does external work; we observe that this model can be used to account for difficulties reported by Atzler & Herbst [20] for the execution of normal walking gaits doing larger amounts of external mechanical work.

## 2 Mechanical and Metabolic Energies of a Movement

We begin with the first part of this project by developing the framework for the description of human movements that we use as the basis of models of limits on the ranges of movements. We first provide a summary of the general segment model that we used in [8]. We next provide a summary of the model for calculating the metabolic energy that we developed in [8]. We then show how to account for the total mechanical energy of a movement by combining the total of the kinetic mechanical energies of the segments of the body with an unspecified mechanism for the storage of mechanical energy in the form of potential mechanical energy. Finally, we look in more detail at two particular forms of mechanical energy that are important in the subsequent sections — external mechanical work and mechanical energy loss — and show how the various energies relate to each other.

*2.1 Segment Model*

We follow [8, 22] in modeling the human body executing movements using a skeleton consisting of a system of N ("nu") segments attached at $N$ joints. As an example, the segment model given in [23] consists of 14 segments: (i) head, (ii) trunk, (iii) upper arms, (iv) lower arms, (v) hands, (vi) thighs, (vii) legs, and (viii) feet, and 18 joints including; (i) 13 points where segments meet, and (ii) 5 segment-endpoints (i.e. the top of the head, the finger-ends of the two hands, and the toe-ends of the two feet). We can specify a movement by having the joints move with joint-trajectories so that the *n-th* joint moves with joint-trajectory $\vec{x}_n(t)$. We denote the velocities of the joints by $\vec{v}_n(t) = \dot{\vec{x}}_n(t)$, and the net effect of the muscle forces by $\vec{F}_n(t)$.

*2.2 Metabolic Energy*

We again follow [8] in developing a framework for modeling the metabolic energy of a movement. We decompose the metabolic rate $\dot{W}(t)$ of expending metabolic energy as the sum of metabolic rates $\dot{W}_n(t)$ associated with each joint $\dot{W}(t) \approx \sum_{n=1}^{N} \dot{W}_n(t)$. The metabolic rate associated with a joint is given by a function:

$$\dot{W}_n(t) = \dot{W}_n^F\left(\vec{F}_n(t)\right) + \dot{W}_n^E\left(\vec{F}_n(t), \vec{v}_n(t)\right). \tag{1}$$

The metabolic rates $\dot{W}_n^F(t)$ and $\dot{W}_n^E(t)$ have the mathematical forms:

$$\begin{aligned} \dot{W}_n^F\left(\vec{F}_n(t)\right) &\approx \varepsilon_n F_n(t)^2, \\ \dot{W}_n^E\left(\vec{F}_n(t), \vec{v}_n(t)\right) &\approx \eta_n \vec{F}_n(t) \cdot \vec{v}_n(t). \end{aligned} \tag{2}$$

The quantities $\varepsilon_n$ and $\eta_n$ are constant parameter values characterizing the associated metabolic rates. We allow that the parameters $\eta_n$ may take on different values when the muscles add or remove mechanical energy to or from a segment, though we require they be constant in each case.

*2.3 Mechanical Energy*

The mechanical energy $U(t)$ is the sum of the mechanical energy $U_{movement}(t)$ required to cause the body to assume the sequence of poses that make up the movement and any mechanical energy $U_{environment}(t)$ the body exchanges with external objects or its environment. We calculate the total kinetic mechanical energy $K_{movement}(t)$ of the body by applying the principles of classical mechanics to the motions of the segments of the model using the trajectories $\vec{x}_n(t)$ of the joints. The potential mechanical energy $V_{movement}(t)$ of the body represents the total effect of any mechanisms the body can use to store and release mechanical energy during a movement. In practice, $V_{movement}(t)$ is a bookkeeping device that we use to provide an explicit account for all the mechanical energy in a movement when using a possibly insufficiently detailed model to represent the movement. It is possible that the energy storage represented by $V_{movement}(t)$ in the model may correspond to movements in space of parts of the physical body not accounted for in the model being used.

*2.4 External Mechanical Work*

The *external mechanical work* $U_{ext}(t)$ is mechanical energy purposefully given to or taken from objects in the course carrying out a specific task (e.g. moving an object). This is mechanical energy the objects require to complete this task, and not mechanical energy incidentally lost to the environment as part of the task (this mechanical energy loss is dealt with in Sec. 2.5). We denote this as *external* mechanical work to emphasize the effect on the environment and to distinguish it from the *internal* mechanical work done to add or remove mechanical energy from the body segments in order for the body to carry out the sequence of poses that comprise the movement.

*2.5 Mechanical Energy Loss*

The *mechanical energy loss* $U_{loss}(t)$ is mechanical energy lost to the environment as part of executing a movement but does not directly contribute to a carrying out a specific task (i.e. it is not a component of the external mechanical work). The mechanical energy loss arises through two mechanisms: (i) by using muscles and expend metabolic energy $\eta \vec{F}(t) \cdot \vec{v}(t)$ to remove mechanical energy from the body segments (i.e. $\vec{F}(t) \cdot \vec{v}(t) < 0$), and (ii) by losing mechanical energy to the environment through an interaction with it (e.g. at heel strike during normal walking gait). As the model that we developed for normal walking gaits in [8] and which we continue to use in this paper does not include any mechanical energy losses using the first mechanism, in this paper, we restrict the term $U_{loss}(t)$ to refer to losses using the second mechanism.

*2.6 Total Rate of Change of Mechanical Energy*

The total rate of change $\dot{K}_{movement}(t)$ of the total kinetic mechanical energy is the sum of the rate $\dot{U}_{metabolic}(t)$ at which the muscles add or remove mechanical energy to or from the body, the rate $\dot{U}_{ext}(t)$ at which the movement does external mechanical work, the rate $\dot{U}_{loss}(t)$ of mechanical energy loss, and the rate $\dot{V}_{movement}(t)$ at which the movement adds or removes mechanical energy to or from a store of potential mechanical energy; we therefore find:

$$\dot{K}_{movement}(t) = \dot{U}_{metabolic}(t) + \dot{U}_{ext}(t) \\ + \dot{U}_{loss}(t) + \dot{V}_{movement}(t). \tag{3}$$

**3 Normal Walking Gaits Doing External Mechanical Work**

We continue to the second part of the project where we develop the model of normal walking gaits doing external mechanical work that we use in the subsequent sections (Secs. 4 and 5) as the basis for the construction of limits on the lowest avg. walking speed and longest avg. step length. In this section, we develop the metabolic and mechanical energies models for normal walking gaits doing external mechanical work while paying particular attention to external mechanical work and mechanical energy loss. We proceed as follows: (i) we give a kinematic model of normal walking gait developed in [8] (Sec. 3.2), (ii) we extend the metabolic energy model for normal walking gaits developed in [8] to develop a model for normal walking gait doing external mechanical work (Sec. 3.3), (iii) we use available empirical data in Atzler & Herbst [20] for a subject walking while pulling a cart to estimate values for the free parameters in the metabolic energy model (Sec. 3.4), (iv) we develop a mechanical energy model of for normal walking gaits that is consistent with the kinematic metabolic energy models that we have developed (Sec.

3.5), and (v) we use the estimated parameter values for the metabolic energy model to estimate the mechanical energy loss during normal walking gaits (Sec. 3.6).

We characterized normal walking gaits in [8] as those middling speed walking gaits selected by healthy young adults and adults under typical conditions and not the very slow walking gaits exhibited by persons with mobility problems or very fast walking gaits with speeds in the region where running becomes preferable. We find it contributes to the clarity of how the metabolic and mechanical energies relate to express metabolic energy in calories and mechanical energy in joules; these two measures of energy are related to each other by 1.0 cal = 4.2 J.

*3.1 Some Anthropometric Values*

For convenience, we give here, in one place, a number of relevant anthropometric values. A subject with mass $M$ and height $H$ has a mass in each leg (i.e. thigh, shank, and foot) of about $\mu = rM$, and the length of the leg of about $L = \rho H$ where $r = 0.16$ and $\rho = 0.53$. [23] The mass of the torso carried by the stance leg during a step is $m = (1 - 2r)M$. The avg. walking speeds and step lengths for adults aged 20-49 years are $v° \approx 1.3$ m·s$^{-1}$ and $s° \approx 0.61$ m. [6] The gross metabolic rate during standing is about 0.31 cal·kg$^{-1}$·s$^{-1}$, and during walking it becomes about 0.50 to 1.7 cal·kg$^{-1}$·s$^{-1}$ for walking speeds of 0.2 m·s$^{-1}$ to 1.9 m·s$^{-1}$. [24]

*3.2 Kinematic Model*

We use the kinematic model for normal walking gaits developed in [8]. This is a two segment model with one segment for each leg − the legs are straight and do not bend at the knee. The mass of the torso is located at the *torso* which is the point where the two segments meet; the mass of each leg is located in the feet at the far end of the leg segments from the torso. The model therefore has three joints: (i) torso, (ii) left foot, and (iii) right foot. We require that the torso maintain a constant height throughout the walk and maintain a constant speed along a straight line parallel to the ground. During one step, one leg is the *stance leg* which supports the torso as the torso moves over it, and the other is the *swing leg* which swings under the torso; the feet of the two legs are the *stance foot* and *swing foot*, respectively. The stance foot remains fixed on the ground while the swing foot glides a negligible distance above the ground; the legs lengthen or shorten as needed by the movement − we intend the amount of lengthening and shortening to be consistent with reasonable knee-bending during walking.

We only look at steady state walking gaits that are in progress and maintain constant values for the gait parameters; we do not look at the process of starting or stopping a walking gait. We describe walking gait using two gait parameters: (i) the avg. walking speed (the constant speed of the torso) $v$, and (ii) the avg. step length (the distance between the feet when they are both on the ground) $s$. We define the unit vector $\hat{v}$ to be the direction of motion of the torso. The gait parameter $v$ giving the walking speed should not be confused with the velocities $\vec{v}_n$ appearing in the metabolic energy model or the unit vector $\hat{v}$; the diacritical mark or the absence of one suffices to distinguish them.

We require the torso to maintain a constant mechanical energy and move in a constant direction horizontally. The effect of this requirement is to make the torso move horizontally with constant speed. We use the model developed in [8] to describe the motion of the swing leg in which the swing leg moves symmetrically so that the swing foot glides horizontally, a negligible distance above the ground, with a constant acceleration during the first half of the step and a constant deceleration during the second half. Mathematically, the motions of the torso and swing foot are:

$$\dot{\vec{x}}_{torso}(t) = v\hat{v},$$
$$\ddot{\vec{x}}_{foot}(t) \approx \begin{cases} (8v^2/s)\hat{v}, & 0 \leq t \leq s/2v, \\ -(8v^2/s)\hat{v}, & s/2v < t \leq s/v. \end{cases} \quad (4)$$

*3.3 Metabolic Energy Model*

We extend the metabolic energy model for normal walking gaits developed in [8] to develop a metabolic energy model for normal walking gaits doing external work. We assume that some amount of mechanical energy has been conserved during walking and that the metabolic energy associated with holding the body up against gravity is approximately constant over the range of walking gaits and is the same as that for standing so that we can treat it as part of the background metabolic rate that we ignore in the calculation of the metabolic energy of a step. We denote the muscle force applied by the stance leg to the torso by $\vec{F}'_{st}(t)$ and the force applied to the swing leg by $\vec{F}_{sw}(t)$; we assume the stance foot is fixed on the ground during the step. We define the external force $F_{ext}$ to oppose the horizontal motion of the torso so that it can be written $-F_{ext}\hat{v}$. The force the body must apply to compensate for the external force is $F_{ext}\hat{v}$. We find that $\vec{F}'_{st}(t) = \vec{F}_{st}(t) + F_{ext}\hat{v}$ where $\vec{F}_{st}(t)$ is the force of the stance leg on the torso when there is no external force. We associate parameters $\varepsilon_{st}$ and $\eta_{st}$ with the stance leg and a parameter $\varepsilon_{sw}$ with the swing leg; these parameters correspond to the parameters in (2). The time required to execute a step is $T = s/v$. The metabolic energy per step is the sum of a constant term $W_0$, and three metabolic energies: (i) the energy expended generating the force $\vec{F}'_{st}(t)$ of the stance leg, (ii) the energy expended generating the force $\vec{F}_{sw}(t)$; of the swing leg, and (iii) the energy expended by the stance leg to provide the mechanical energy of the external mechanical work; this is:

$$W \approx W_0 + \varepsilon_{st}\int_0^{s/v} F'_{st}(t)^2 \, dt + \varepsilon_{sw}\int_0^{s/v} F_{sw}(t)^2 \, dt + \eta_{st}F_{ext}v\int_0^{s/v} dt. \quad (5)$$

We expand $F'_{st}(t)^2 = F_{st}(t)^2 + 2F_{ext}\vec{F}_{st}(t)\cdot\hat{v} + F_{ext}^2$ in (5), and we find:

$$W \approx W_0 + \varepsilon_{st}\int_0^{s/v} F_{st}(t)^2 \, dt + \varepsilon_{sw}\int_0^{s/v} F_{sw}(t)^2 \, dt$$
$$+ 2\varepsilon_{st}F_{ext}\int_0^{s/v} \vec{F}_{st}(t)\cdot\hat{v}\,dt + (\varepsilon_{st}F_{ext}^2 + \eta_{st}F_{ext}v)\int_0^{s/v} dt. \quad (6)$$

The values of the parameters $W_0$, $\varepsilon_{st}$, $\varepsilon_{sw}$, and $\eta_{st}$ must be estimated empirically.

We may write (5) as the sum $W = W_{gait} + W_{ext}$ of: (i) the metabolic energy $W_{gait}$ per step needed to move the segments of the body through the sequence of poses comprising one step and compensate for mechanical energy lost by the body (this is the metabolic energy for normal walking gaits calculated in [8]), and (ii) the metabolic energy $W_{ext}$ per step above $W_{gait}$ that gives the additional metabolic energy the body must expend above that needed for normal walking gait in order to do the external mechanical work; we find:

$$W_{gait} \approx W_0 + \varepsilon_{st}\int_0^{s/v} F_{st}(t)^2 \, dt + \varepsilon_{sw}\int_0^{s/v} F_{sw}(t)^2 \, dt,$$
$$W_{ext} \approx 2\varepsilon_{st}F_{ext}\int_0^{s/v} \vec{F}_{st}(t)\cdot\hat{v}\,dt + (\varepsilon_{st}F_{ext}^2 + \eta_{st}F_{ext}v)\int_0^{s/v} dt. \quad (7)$$

We look first at the muscle forces $\vec{F}_{st}(t)$ and $F_{ext}$ applied by the stance leg to the torso as illustrated in Fig 1. Again, $\vec{F}_{st}(t)$ is force the stance leg must produce to move the torso as required and $F_{ext}$ the further force the stance leg must produce to perform the external work so $\vec{F}'_{st}(t) = \vec{F}_{st}(t) + F_{ext}\hat{v}$. In (4), we require the torso to move horizontally with constant velocity $\dot{\vec{x}}_{torso}(t)$. During the first half of the step, gravity pulls the torso in the direction $-\hat{v}$, while, during the second half, gravity pulls the torso in the direction $\hat{v}$. The torso already has the mechanical energy needed to carry it forward with the required speed, so the muscle forces counteract the force of gravity and do external mechanical work.

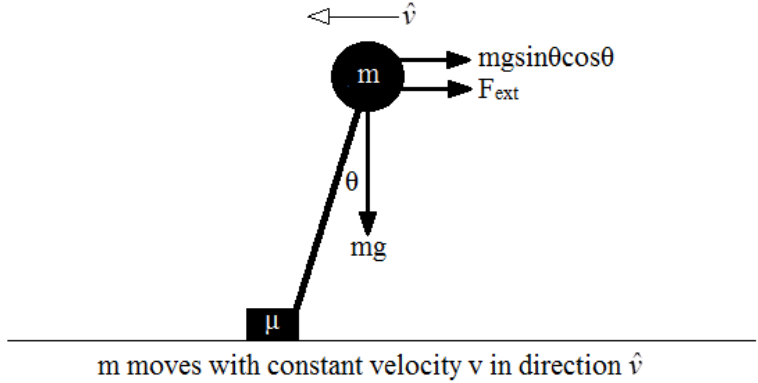

**Figure 1.** The motion of the torso over the stance leg during one step of walking gait. The torso moves over the leg at a constant speed $v$ while producing and external force $F_{ext}$. To maintain the constant speed, the muscles must provide force to compensate for the effect of gravity on the torso's speed.

The angle $\theta(t)$ of the stance leg with respect to the vertical determines the effect of gravity on the torso. We define $\theta(t)$ so that it is negative during the first half of the step, and positive during the second half. The horizontal force related to the torque on the inverted pendulum by gravity is $mg\sin\theta\cos\theta = (1/2)mg\sin 2\theta$ where $g$ denotes the acceleration of gravity; we find that $\theta(t)$, $\vec{F}_{st}(t)$, and $\vec{F}'_{st}(t)$ satisfy:

$$
\begin{aligned}
&L\sin\theta(t) = vt - s/2, \\
&\vec{F}_{st}(t) = (1/2)mg\sin 2\theta(t)\hat{v}, \\
&\vec{F}'_{st}(t) = (1/2)mg\sin 2\theta(t)\hat{v} + F_{ext}\hat{v}.
\end{aligned}
\quad (8)
$$

We look next at the muscle forces $\vec{F}_{sw}(t)$ applied by the torso to the swing leg as illustrated in Fig. 2. We require the swing foot to move horizontally with acceleration $\ddot{\vec{x}}_{foot}(t)$ in (4). During one step, the body moves a distance of one avg. step length, and the swing foot travels a distance of one stride length horizontally in direction $\hat{v}$. The swing leg begins and ends the swing at rest, and so must accelerate and decelerate as required over the course of the swing. During the first half of the step, gravity pulls the swing leg in the direction $\hat{v}$, while, during the second half, gravity pulls the swing leg in the direction $-\hat{v}$.

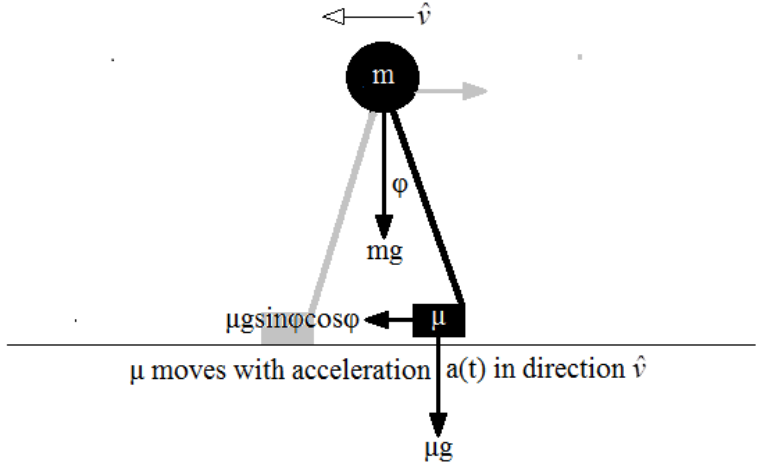

**Figure 2.** The motion of the swing leg under the torso during one step of walking gait. The swing leg moves symmetrically under the torso, accelerating with an acceleration $a(t)$ during the first half of the step, and decelerating with an acceleration $a(t)$ during the second half. The muscles must provide the force that generates the acceleration and deceleration and compensate for the effect of gravity on the swing leg.

The angle $\varphi(t)$ of the swing leg with respect to the vertical determines the effect of gravity on the swing leg. We define $\varphi(t)$ so that it is negative during the first half of the step, and positive during the second half. As the acceleration of the swing leg is symmetric through the step, we can calculate the metabolic energy of the swing leg by doubling the metabolic energy for the first half of the step. The horizontal muscle forces that must be applied to the leg to generate the constant acceleration in (4) during the first half of the step is $\mu a + \mu g \sin\varphi\cos\varphi = \mu a + (1/2)\mu g \sin 2\varphi$; we find that $\varphi(t)$ and $\vec{F}_{sw}(t)$ for the first half of the step satisfy:

$$L\sin\varphi(t) = \left(4v^2/s\right)t^2 - vt - s/2, \quad 0 \le t \le s/2v,$$
$$\vec{F}_{sw}(t) = \left(\left(4\mu v^2/s\right) + \left(\mu g/2\right)\sin 2\varphi(t)\right)\hat{v}. \tag{9}$$

Combining (7), (8), and (9), and using the observation that for small avg. step lengths $s$, we may take $\sin 2\theta \approx 2\sin\theta$ and $\sin 2\varphi \approx 2\sin\varphi$, we calculate the metabolic energy $W_{gait}$ per step to be:

$$W_{gait} \approx W_0 + \left(\varepsilon_{st} m^2 g^2 / 12L^2\right)s^3/v + \left(\varepsilon_{sw}\mu^2 g^2/5L^2\right)s^3/v \\ - \left(10\varepsilon_{sw}\mu^2 g/3L\right)vs + \left(16\varepsilon_{sw}\mu^2\right)v^3/s. \tag{10}$$

Combining (7) and (8), and noting that the first term on the right-hand side of (7) is a symmetric integral of an odd function and therefore zero, we calculate the metabolic energy $W_{ext}$ per step to be:

$$W_{ext} \approx \left(\varepsilon_{st}\right)F_{ext}^2 s/v + \left(\eta_{st}\right)F_{ext}s. \tag{11}$$

*3.4 Empirical Study (Atzler & Herbst, 1927)*

We now make the metabolic energy model developed in (5), (7), (10), and (11) into an estimator of the metabolic energy during walking over the range of allowed walking gaits by using empirical data to produce estimates for the values of the parameters $W_0$, $\varepsilon_{st}$, $\varepsilon_{sw}$, and $\eta_{st}$. The data set we use includes the data we used in [8] to estimate the parameter values $W_0$, $\varepsilon_{st}$, and $\varepsilon_{sw}$ for the normal walking gaits model and also includes data for the same subject where external mechanical work was done during normal walking gait.

Atzler & Herbst [20, 25] observed one subject (male, $M = 68$ kg, $H = 1.7$ m, aged 39 years, mass-normalized resting metabolic rate $\dot{w}_{rest} = \dot{W}_{rest}/M = 0.30$ cal·kg$^{-1}$·s$^{-1}$) perform a variety walking gaits, and measured the metabolic energy for each walking gait using a Zuntz-Geppert respiratory apparatus and the Douglas Bag Technique. The subject was trained to walk on a horizontal treadmill using all 20 combinations of 4 avg. step lengths $s$ (0.46 m, 0.60 m, 0.76 m, and 0.90 m) and 5 avg. cadences $v/s$ (0.83 step·s$^{-1}$, 1.25 step·s$^{-1}$, 1.7 step·s$^{-1}$, 2.2 step·s$^{-1}$, and 2.5 step·s$^{-1}$). In addition to walking freely, the subject walked pulling a cart (Deichselwagen) such that the subject had to apply four different external forces $F_{ext}$ to overcome friction (100 N, 110 N, 130 N, and 160 N). The handle by which the subject pulled the cart attached to the cart via a rigid shaft that was fixed so that the handle was positioned 1.0 m above the floor. For the setups requiring external forces $F_{ext}$ of 100 N and 110 N, the subject performed walking gaits at all 20 combinations of step length and cadence. The subject experienced "disagreeable difficulties in breathing" when pulling the heavier carts at the higher walking speeds for the required time, so some combinations of walking speed, cadence, and external force were left out of the study. For the setups requiring external force $F_{ext} = 130$ N, the subject performed 18 of the 20 combinations, leaving out walking gaits with avg. step lengths and avg. cadences of (i) $s = 0.90$ m and $v/s = 2.2$ step·s$^{-1}$, and (ii) $s = 0.90$ m and $v/s = 2.5$ step·s$^{-1}$ For the setups requiring external force $F_{ext} = 160$ N, the subject performed 12 of the 20 combinations, leaving out walking gaits with avg. step length of 0.90 m or avg. cadence 0.83 step·s$^{-1}$. The fact that the subject declined to perform some of the scheduled normal walking doing external work tasks also suggests the tasks with the large external forces are nearing a limit on the range of movement for the subject; we propose a limit to account for this in Sec. 5.

We estimated the parameter values $W_0$, $\varepsilon_{st}$, $\varepsilon_{sw}$, and $\eta_{st}$ using only the data for $F_{ext}$ of 0 N, 100 N, 110 N, and 130 N. Although the model still worked fairly well when the data for an external force of 160 N were included, it appeared the model was beginning to perform insufficiently well when applied to these large external forces. We discuss this further in Sec. 3.7, and propose a model for the inability to walk with avg. step length of 0.90 m for $F_{ext} = 160$ N in Sec. 5.

Combining (5), (7), (10), and (11), we find:

$$W \approx W_0 + \left(m^2g^2s^3 / 12L^2v + F_{ext}^2 s / v\right)\varepsilon_{st} \\ + \left(16\mu^2 v^3 / s - 10\mu^2 gvs / 3L + \mu^2 g^2 s^3 / 5L^2 v\right)\varepsilon_{sw} + \left(F_{ext}s\right)\eta_{st}. \tag{12}$$

Using ordinary least-squares regression, this model fit the data for Atzler & Herbst's subject with $R^2 = 0.96$ and p < 0.0001, and the parameter values were:

$$W_0 \approx 7.0\,cal,$$
$$\varepsilon_{st} \approx 1.9 \times 10^{-3}\,cal \cdot N^{-2} \cdot s^{-1},$$
$$\varepsilon_{sw} \approx 2.6 \times 10^{-3}\,cal \cdot N^{-2} \cdot s^{-1},$$
$$\eta_{st} \approx 0.62\,cal \cdot J^{-1}. \tag{13}$$

Inspection of the 95% confidence intervals showed that all 4 parameters were statistically significant. The fit is illustrated in Fig. 3 using all the walks observed by Atzler & Herbst. We chose to include walks with external force $F_{ext} = 160$ N to provide an indication of the extent to which the model begins to break down for large external forces. The estimated parameter values in (13) are close to the values of $W_0 \approx 9.0$ cal, $\varepsilon_{st} \approx 2.5 \times 10^{-3}$ cal $\cdot$ N$^{-2}$ $\cdot$ s$^{-1}$, and $\varepsilon_{sw} \approx 1.7 \times 10^{-3}$ cal $\cdot$ N$^{-2}$ $\cdot$ s$^{-1}$ estimated in [8] where the same data were analyzed only for the case where $F_{ext} = 0$.

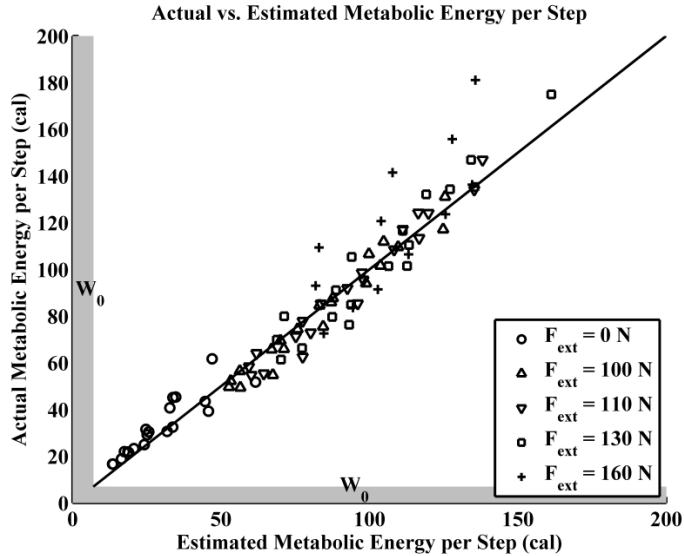

**Figure 3.** Actual vs. estimated metabolic energy per step. We use the model in (12) to fit the observed metabolic energies for the 78 walking gaits in Atzler & Herbst with external forces $F_{ext} = 0$ N, 100 N, 110 N, and 130 N; the fit for the model for these walking gaits has R$^2$ = 0.96 and p < 0.0001. We have plotted the data for all 90 walking gaits in Atzler & Herbst, including walking gaits with $F_{ext} = 160$ N using the fitted model; the value of $F_{ext}$ is indicated for each data point. We have included walking gaits with $F_{ext} = 160$ N to provide an indication of how much the model breaks down for large $F_{ext}$. The value for the constant parameter W$_0$ is indicated. For reference, we show a segment of the line with slope one passing through the origin.

*3.5 Mechanical Energy Model for Normal Walking Gaits*

As we defined the movement for normal walking gaits in (4), the torso maintains a constant kinetic mechanical energy while the kinetic mechanical energy of the swing leg varies through the step. We allow for a mechanism where some of the kinetic mechanical energy of the swing leg can be stored as a potential mechanical energy $V_{sw}(t)$ and used to produce kinetic mechanical energy in the other leg when it becomes the swing leg. We assume that normal walking gaits bear some resemblance to passive-dynamic walking models [26, 27] and allow some kinetic mechanical energy of the swing leg be lost at heel strike. We model this by allowing the kinetic mechanical energy to be lost during a very short time $\delta t$ beginning at each heel strike. We assume that mechanical energy loss during normal walking gaits only occurs at heel strikes, so with each heel strike the body loses a mechanical energy $U_{loss}$; since we have required the torso to maintain a constant mechanical energy, this energy loss must come from the swing leg. After toe-off, the swing foot accumulates kinetic mechanical energy until it reaches a maximum at midswing after which the kinetic mechanical energy of the swing foot falls back to zero at heel strike. We assume that after midswing, the body stores the kinetic mechanical energy that the swing foot is losing using a potential mechanical energy $V_{sw}(t)$. At heel strike the swing foot has no kinetic mechanical energy, so we take $U_{loss}$ entirely from the potential mechanical energy $V_{sw}(t)$; the body recycles whatever mechanical energy remains in $V_{sw}(t)$ after the loss of $U_{loss}$ into the swing of the leg in the next step and restores the lost mechanical energy using the muscles.

*3.6 Mechanical Energy Loss for Normal Walking Gaits*

The model in (12) includes a constant term $W_0$ representing an additional contribution of metabolic energy for each step that is not accounted for in the metabolic energy model developed in Sec. 3.4. $W_0$ is not due to the resting metabolic rate as Atzler & Herbst have factored this out of their results and reported the metabolic energies above resting. As the metabolic energy model accounts for all the forces needed to make the body walk, the only reason for an additional expenditure of metabolic energy $W_0$ would be for recovering the mechanical energy $U_{loss}$ lost each step. Since walking consists of a sequence of steps, the lost mechanical energy must be restored to the swing leg each step by the action of muscles on the swing leg; we therefore find $U_{loss} = W_0 / \eta_{sw}$. Since $\varepsilon_{sw} \approx \varepsilon_{st}$ in (13) we assume that $\eta_{sw} \approx \eta_{st}$, and estimate the mechanical energy loss to be:

$$U_{loss} \approx W_0 / \eta_{st}. \tag{14}$$

Using the parameter values in (13), we find:

$$U_{loss} \approx 11 J. \tag{15}$$

Using a typical walking cadence of $v°/s° \approx 2.1$ step·s$^{-1}$, we can estimate that a typical walking gait loses mechanical energy at an avg. rate of about $\langle \dot{U}_{loss} \rangle \approx 23$ W. Researchers have developed a variety of techniques for harvesting energy from the walking human body while minimally affecting the subject's normal walking gait. Some energy harvesting techniques yielding larger powers include: knee-mounted energy harvester yielding 5 W, [28] and suspended-load backpack yielding 7.4 W, [29] both well below the estimated $\langle \dot{U}_{loss} \rangle$. The power generated by these techniques lie well below the avg. rate of mechanical energy loss that we have estimated.

*3.7 Discussion*

The model for the metabolic energy of normal walking gaits doing external mechanical work that we have developed in this section decomposes into the sum of the metabolic energy $W_{gait}$ needed to move the body and the metabolic energy $W_{ext}$ needed to do the external mechanical work. As we have seen in Sec.

3.4, this model for $W_{ext}$ breaks down for large external forces $F_{ext}$. In [8], we derived the formulas in (2) for calculating metabolic work by approximating $\dot{W}_n^F\left(\vec{F}_n(t)\right)$ and $\dot{W}_n^E\left(\vec{F}_n(t), \vec{v}_n(t)\right)$ using Taylor series truncated to the lowest order non-zero term. The breakdown of the model for $W_{ext}$ in the case of large external forces $F_{ext}$ suggests that while the approximation we used in [8] works for smaller muscle forces, it may need to be modified to properly account for larger muscle forces either by truncating the Taylor series at higher order terms, or by introducing terms that are functions of variables other than the $\vec{v}_n(t)$ or $\vec{F}_n(t)$.

The mechanical energy $U_{loss}$ lost each step that we have estimated is that for a normal middle-aged adult male. We expect that the mechanical energy loss for individuals may vary from this value for any number of reasons. However, on particular reason of interest to the present discussion is that of an individual exhibiting some sort of motor control deficit affecting walking gaits. In this case, as we expect such an individual's walking gait to fail to exhibit the nuanced control and coordination characteristic of normal healthy walking gaits, the normal walking gait should present a higher mechanical energy loss.

## 4 Lower Limits on Avg. Walking Speed & Very Slow Walking Gaits

We now more on to the third part of the project where we use the model for normal walking gaits doing external work developed in Sec. 3 as the basis for constructing models for lower limits on the avg. walking speeds of walking gaits. We build these models by providing an account of how the mechanical energy loss $U_{loss}$ per step is taken from the mechanical energy of the body. We arrive at two lower limits and define the range between the limits as that of very slow walking gaits. We observe that the slowest avg. walking speeds reported for younger adults by Grieve, [19] for older adults by Hagler et al., [21] for pre-rehabilitation PD subjects in Frazzitta et al. [7] lie around the range of very slow walking gaits defined by these two limits.

*4.1 Lower Limit on Avg. Walking Speed for Normal Walking Gaits*

In the mechanical energy model that we have developed in Sec. 3.5, we required that the torso maintain a constant mechanical energy and that the mechanical energy loss $U_{loss}$ come entirely from the mechanical energy of the swing leg. The kinetic mechanical energy of the swing foot attains a maximum value of $2\mu v^2$ and falls to zero at heel strike as this mechanical energy is stored as a potential mechanical energy $V_{sw}$. It follows that we must have $V_{sw} \geq U_{loss}$, and that normal walking gaits have a lower limit at $V_{sw} = U_{loss}$. Thus, the lower limit occurs at an avg. walking speed $v_*$ where $2\mu v_*^2 = U_{loss}$; that is:

$$v_*\left(U_{loss}\right) = \sqrt{\left(1/2r\right)\left(U_{loss}/M\right)}. \tag{16}$$

As we indicate explicitly in (16), the limit $v_*(U_{loss})$ depends on the mechanical energy loss; we discuss this further in Sec. 4.4.

*4.2 Very Slow Walking Gaits Model*

A subject can walk still slower than the limit $v_*(U_{loss})$ by allowing for loss of mechanical energy from the torso at heel strike. We can modify mechanical energy model for normal walking gaits in Sec. 3.5 to produce a very slow walking gait model that accommodates very slow avg. walking speeds that lie below $v_*(U_{loss})$ by taking $U_{loss}$ from all the mechanical energy available in $V_{sw}$ with the remainder coming from the torso. In this case, we require the torso to lose and restore the required mechanical energy during a very short time $\delta t$ following heel strike. We model this process by introducing a rudimentary double support phase to the gait cycle and expand toe-off/heel strike so that a very short time interval $\delta t$ occurs between the heel strike of one leg and the toe-off of the next. This allows the required changes in the mechanical energy of the torso to occur during the interval $\delta t$, while the mechanics of a swing of the foot approximate the model that we have developed in Sec. 3. In this model, the muscles acting on the swing leg restore the mechanical energy lost by the swing leg through the parameter $\eta_{sw}$ while the muscles acting on the torso restore the mechanical energy lost by the torso through the parameter $\eta_{st}$. As we have assumed $\eta_{sw} \approx \eta_{st}$, (14) still holds to describe the mechanical energy loss during very slow walking gaits. The torso and the swing leg have enough mechanical energy to supply the needed mechanical energy loss $U_{loss}$ for each step so long as $V_{sw} + (1/2)mv^2 \geq U_{loss}$, so very slow walking gaits have a lower limit of $V_{sw} + (1/2)mv^2 = U_{loss}$. Again, the maximum kinetic mechanical energy of the swing foot is $2\mu v^2$, so the lower limit occurs at an avg. walking speed $v_{min}$ where $(1/2)(m + 4\mu)v_{min}^2 = U_{loss}$; that is:

$$v_{\min}\left(U_{loss}\right) = \sqrt{\left(2 / \left(1 + 2r\right)\right)\left(U_{loss} / M\right)}. \tag{17}$$

As we indicate explicitly in (17), the limit $v_{min}(U_{loss})$ depends on the mechanical energy loss; we discuss this further in Sec. 4.4. The limit $v_{min}(U_{loss})$ does not depend on the model of mechanical energy loss that we are using; we expect it to hold generally.

We illustrate the limits on normal and very slow walking gaits in Fig. 4. We have used the height-normalized avg. stride length rather than the avg. step length in Fig. 4 to facilitate comparison to the corresponding figure in [8]. The limits corresponding to $v_*/H$ and $v_{min}/H$ for Atzler & Herbst's subject are indicated by vertical lines. The limit SP corresponds to a walking gait where the swing leg moves as a pendulum under the force of gravity; we have argued in [8] that this should constitute a boundary on feasible walking gaits. The range of normal walking gaits is indicated in white, the range of very slow walking gaits is indicated in light grey, and the region where the normal and very slow walking gait models do not apply in darker grey. The typical adult normal walking gait is indicated with height-normalized avg. walking speed v°/H and avg. stride length 2s°/H.

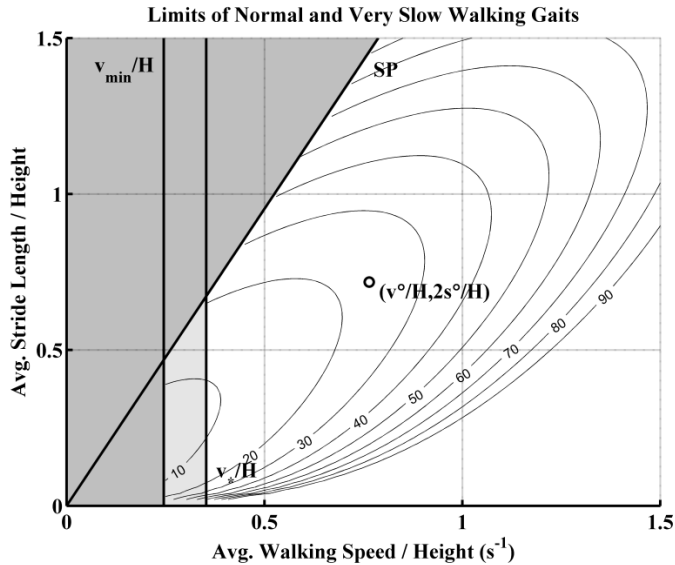

**Figure 4.** Limits of normal and very slow walking gaits. The limits corresponding to $v_*/H$ and $v_{min}/H$ are indicated by vertical lines. The limit SP corresponds to a walking gait where the swing leg moves as a pendulum under the force of gravity. The range of normal walking gaits is indicated in white, the range of very slow walking gaits is indicated in light grey, and the region where the normal and very slow walking gait models do not apply in darker grey. The typical adult normal walking gait is indicated with height-normalized avg. walking speed v°/$H$ and avg. stride length 2s°/$H$. The contour lines correspond indicate constant metabolic energies per step (in calories) in the metabolic energy landscape. The typical adult walking gait is indicated with height-normalized avg. walking speed $v°/H$ and avg. stride length $2s°/H$.

*4.3 Empirical Studies*

We estimate the avg. walking speeds $v_*$ and $v_{min}$ for Atzler & Herbst's subject to be $v_* \approx 0.72$ m · s⁻¹ and $v_{min} \approx 0.50$ m · s⁻¹, or $v_*/H \approx 0.42$ step · s⁻¹ and $v_{min}/H \approx 0.29$ step · s⁻¹ when height-normalized. We can compare these values to observations of subjects in Grieve. [19] Grieve observed young adult subjects perform a variety walks and measured gait parameter values for the walks from film, using an analyzing projector. The subjects were prompted to walk at a variety of avg. walking speeds using instructions like "very slowly and relaxed," "a little faster," and so on until they were walking as fast as possible after about ten transits. The lowest height-normalized avg. walking speeds $v/H$ executed by the subjects in Grieve were about $v/H \approx 0.25$ step · s⁻¹, so slightly below the $v_{min}$ that we have estimated. We can also compare these values to observations of subjects in Hagler et al. [21] Hagler et al. observed older adult subjects perform a variety of walks and measured avg. walking speed using a GAITRite® Walking System gait mat. The subjects were prompted to walk at a variety of avg. walking speeds described as "slow," "normal," and "fast." The lowest avg. walking speeds $v$ executed by the subjects in Hagler et al. were about $v \approx 0.40$ m · s⁻¹, or $v/H \approx 0.24$ step · s⁻¹ when height-normalized using an avg. height of 1.7 m, so, again, slightly below the $v_{min}$ that we have estimated. We can further compare the values to observed very slow walking gaits of pre-rehabilitation PD subjects reported by Frazzitta et al. [7] Frazzitta et al. reported avg. walking speed both before and after a rehabilitation treatment. The walking gaits of Frazzitta et al.'s subjects before the rehabilitation treatment were reported to have avg. walking speeds of $v \approx 0.50$ m·s$^{-1}$ to 0.60 m·s$^{-1}$, or $v/H \approx 0.29$ step·s$^{-1}$ to 0.35 step·s$^{-1}$ when height-normalized using an avg. height of 1.7 m, so at or slightly above the $v_{min}$ that we have estimated. Thus, before the rehabilitation treatment, the pre-rehabilitation PD subjects appear to have selected walking gaits that have avg. walking speeds lying within lower half of walking speeds in the range of very slow walking gaits that we have identified.

*4.4 Discussion*

We have defined a range of very slow walking gaits between the avg. walking speed limits $v_*$ and $v_{min}$. While more sophisticated models of the mechanical energy loss may affect how we view the region near the limit $v_*$, we expect that $v_*$ will still provide still be convenient in providing a clear means of specifying the slowest walking gaits  The model for normal and very slow walking gaits does not forbid some form of bipedal locomotion with avg. walking speeds below $v_{min}$, however it does not consider these forms of locomotion to be a kind of "walking gait" as they lack the presence of mechanical energy in the torso to carry the body through heel strike into the next step. Since the body comes to a stop with each heel strike, we expect that the step-to-step variability of these forms of locomotion will show a higher irregularity than observed for normal and very slow walking gaits.

We have given the avg. walking speed limits in (16) and (17) as functions $v_*(U_{loss})$ and $v_{min}(U_{loss})$ of the mechanical energy loss $U_{loss}$ per step. Thus, these two lower limits may shift to the right or left in Fig. 4 as $U_{loss}$ changes. Changes in $U_{loss}$ may arise for any number of effects that result in changes in the subject's walking gait, whether due to deliberate changes in how the subject controls the gait or due performance deficits arising from aging or disease. We argued in Sec. 3.7 that performance deficits should affect $U_{loss}$ by increasing the mechanical energy loss per step. We thus find that performance deficits should shift $v_*(U_{loss})$ and $v_{min}(U_{loss})$ in the direction of higher avg. walking speeds. As a result, when walking very slowly, need to walk faster to maintain mechanical energy in the torso that can carry the body in to the next step, or must use the halting form of bipedal locomotion that lies below $v_{min}(U_{loss})$ at higher avg. walking speeds.

**5 Maximum Avg. Step Length Limit**

We conclude with the final part of this project where we use the model for normal walking gaits doing external work developed in Sec. 3 as the basis for constructing a model for the upper limit on the avg. step lengths of walking gaits. Atzler & Herbst observed that their subject experienced "disagreeable difficulties in breathing" for the setups requiring external force $F_{ext} = 130$ N and avg. step lengths and avg. cadences of (i) $s = 0.90$ m and $v/s = 2.2$ step·s$^{-1}$, and (ii) $s = 0.90$ m and $v/s = 2.5$ step·s$^{-1}$, as well as for setups requiring external force $F_{ext} = 160$ N and avg. step lengths of 0.90 m or avg. cadences 0.83 step·s$^{-1}$ (Sec. 3.4). We can account for this by proposing that for large external forces $F_{ext}$, the subject bumps up against one or more limits at which the subject is no longer able to maintain the required level of exertion. The force is that exerted on the torso by the stance leg; one limit that we can place on this force is that it cannot exceed some maximum value $F_{max}$. We may think of this $F_{max}$ as providing a measure of the "strength" of the subject in a similar manner to "strength" metric proposed in [22].

*5.1 Upper Limit on Avg. Step Length for Normal Walking Gaits Doing External Work*

We would like to place an upper limit $F_{max}$ on the amount of force the muscles can generate during normal walking gaits doing external work; we assume that the limit $F_{max}$ is high enough that the body can execute normal walking gaits doing no external work. Since the external force $\mathbf{F}_{ext}$ does not appear in the model for the swing leg in (9), we may therefore assume that the limit does not affect the motion of the swing leg. Therefore, we examine the effect of the upper limit on the stance leg. The total force $\vec{F}'_{st}(t)$ exerted to both move the torso and generate the external force is given in (8). Using the observation that for small avg. step lengths $s$, we may take $sin2\theta \approx 2sin\theta$, we find that it is $max(\vec{F}'_{st}(t)) \approx mgs/2L + |\mathbf{F}_{ext}|$. We require that $max(\vec{F}'_{st}(t)) \leq F_{max}$; therefore, the maximum avg. step lengthy $s_{max}$ is:

$$s_{\max}\left(F_{ext}\right) \approx 2L\left(F_{\max} - \left|F_{ext}\right|\right) / mg. \tag{18}$$

Atzler & Herbst's subject [20, 25] (see Sec. 3.5) was able to carry out tasks with avg. step lengths of 0.90 m when $\mathbf{F}_{ext} = 130$ N, but not $\mathbf{F}_{ext} = 160$ N. We may account for this observation using (18) with an approximate value for $F_{max}$ of:

$$F_{\max} \approx 370N. \tag{19}$$

For comparison, the force the subject's legs exert to hold up the torso against gravity is $mg = 450$ N. The model of the maximum avg. step length limit is illustrated in Fig. 5. The walks observed by Atzler & Herbst for $\mathbf{F}_{ext} = 130$ N and $\mathbf{F}_{ext} = 160$ N are indicated. The maximum avg. step length limit lies above avg. step lengths of 0.90 m when $\mathbf{F}_{ext} = 130$ N, but below those avg. step lengths when $\mathbf{F}_{ext} = 160$ N. We note that the lowest avg. cadence walks for $\mathbf{F}_{ext} = 130$ N lie to the left of (but nevertheless very near to) the limit SP; we discuss this further in Sec. 5.2.

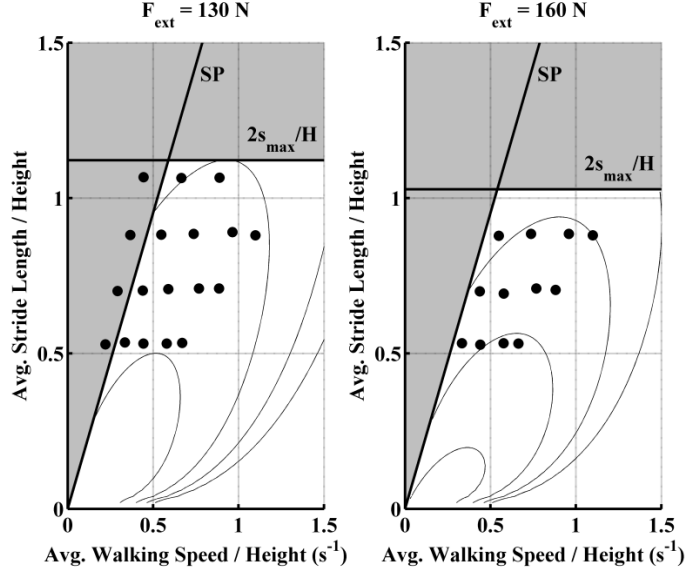

**Figure 5.** Limits of normal walking gaits doing external work. The limits corresponding to $2s_{max}/H$ are indicated by horizontal lines. The limit SP corresponds to a walking gait where the swing leg moves as a pendulum under the force of gravity. The walks at which Atzler & Herbst's subject was able to execute walks given the external force are indicated by black dots. The contour lines correspond indicate constant metabolic energies per step (in calories) in the metabolic energy landscape. We note that the position of the limit corresponding to $2s_{max}/H$ allows the subject to walk with longer avg. step lengths when $F_{ext} = $ 130 N then when $F_{ext} = $ 160 N.

We can estimate an upper limit on the avg. step length of normal walking gaits by solving (19) for the case where $F_{ext} = 0$ N; we estimate a maximum avg. step length limit of 1.5 m. This is less than the subject's height of 1.7 m and gives a height-normalized avg. stride length of $2s_{max}/H = 1.8$ in Fig. 4. This value is well above any height-normalized avg. stride lengths observed by Grieve, [19] and is therefore consistent with those data. This is an upper limit on the force the subject can generate during walking gait; it is possible other limits may exist to lower the maximum avg. step length (e.g. a purely mechanical limit on how the legs can be position during walking gait).

*5.2 Discussion*

The limit we have developed in (19) does not account for the trouble the subject had of walking with $F_{ext} = 130$ N using avg. step lengths and avg. cadences of (i) $s = 0.90$ m and $v/s = 2.2$ step/s, and (ii) $s = 0.90$ m and $v/s = 2.5$ step/s. We should be able to model thus using another limit; however the mode of this limit should be more involved than that developed in this section as it would involve a combination of the avg. walking speed and avg. step length. The limit SP is an approximate limit derived using a simple model of the swing leg so it is possible that a more detailed model of the swing leg might give a corresponding limit in a different position in the avg. walking speed/avg. step length plane. While the subject was able to execute walks to the left of this limit, the subject was unable to execute the walks to the left of this limit for $F_{ext} = 160$ N, suggesting that the limit that SP approximates moved as the external force increased. If the limit that SP approximate also applies to the swing leg, then the observation that it moves as the external force $F_{ext}$ indicates that the external force affects the motion of the swing leg, an effect that is not present in the model that we have developed to describe normal walking gaits doing external work.

We have given the avg. step length limit in (18) as a function $s_{max}(F_{max})$ of the maximum force $F_{max}$ the stance leg can exert on the torso. Thus, the upper limit may shift up or down in Fig. 5 as $F_{max}$ changes. We have presented $F_{max}$ as a kind of measure of "strength," so we expect a decline in $F_{max}$ as a subject becomes weaker. Thus we find that a weakening of subjects due to aging or disease results in subjects typically making shorter steps during normal and very slow walking gaits as one would expect, and due to the high correlation between avg. walking speed and avg. step length, likely a slowing of walking gait as well.

## Acknowledgements

This work was supported by supported by the National Science Foundation under Grants 1407928 and 1111965 and by the National Institute on Aging under Grants NIA P30AG024978 and 5RC1AG36121-2. We thank our colleagues from Northeastern University, Holly Jimison and Misha Pavel, who provided insight and expertise that greatly assisted the research, although they may not agree with all of the interpretations/conclusions of this paper.